\documentclass[usenatbib]{mn2e}
\usepackage{graphicx}

\title[]{Changes in the red giant and dusty environment of the recurrent nova
RS~Ophiuchi following the 2006 eruption}

\author[Rushton et al.]
{M. T. Rushton,$^{1}$ B. Kaminsky,$^2$ D. K. Lynch,$^{3}$\thanks{Visiting 
Astronomer at the Infrared Telescope Facility, which is operated by 
the University of Hawaii under Cooperative Agreement NCC 5-538 with the
National Aeronautics and Space Administration, Science Mission Directorate,  
Planetary Astronomy Program.}, Ya. V. Pavlenko,$^2$, A. Evans,$^4$, \newauthor 
S. P. S. Eyres$^1$, C. E. Woodward$^{5}$\footnotemark[1],
Ray W. Russell$^{3}$\footnotemark[1], Richard J. Rudy$^{3}$\footnotemark[1],
\newauthor Michael L. Sitko$^{6,7}$\footnotemark[1], T. Kerr$^8$ \\
$^1$Jeremiah Horrocks Institute for Astrophysics and Supercomputing, University
of Central Lancashire, Preston, PR1 2HE, UK \\
$^2$Main Observatory of the National Academy of Sciences of Ukraine, 27
Zabolonnoho, Kyiv-127,03680, Ukraine \\
$^3$The Aerospace Corporation, Mail Stop M2-266, P.O. Box 92957, Los Angeles, CA
90009-2957, USA\\ 
$^4$Astrophysics Group, Keele University, Keele, Staffordshire,  ST5 5BG, UK\\
$^5$Department of Astronomy, University of Minnesota, 116 Church Street, S.~E., 
Minneapolis, MN 55455, USA \\
$^6$Space Science Institute, 4750 Walnut Street, Suite 205, Boulder, CO 80301, USA \\
$^7$Department of Physics, University of Cincinnati, Cincinnati, OH 45221, USA
\\
$^8$Joint Astronomy Centre, 660 North A'ohoku Place, University Park, Hilo,
HI 96720, USA
}

\date{Version of 9 Jul 2009}

\pagerange{\pageref{firstpage}--\pageref{lastpage}} \pubyear{2009}

\def\LaTeX{L\kern-.36em\raise.3ex\hbox{a}\kern-.15em
    T\kern-.1667em\lower.7ex\hbox{E}\kern-.125emX}
\let\leqslant=\leq 
\newcommand{\vunit}{\mbox{\,km\,s$^{-1}$}}
\newcommand{\mic}{\mbox{$\,\mu$m}}

\newcommand{\gtsimeq}{\raisebox{-0.6ex}{$\,\stackrel
        {\raisebox{-.2ex}{$\textstyle >$}}{\sim}\,$}}

\begin{document}
\label{firstpage}
\maketitle

\begin{abstract}

We present near infrared spectroscopy of the recurrent nova RS~Oph obtained on
several occasions after its latest outburst in 2006 February. The
$1-5$\mic\ spectra are dominated by the red giant, but the H\,{\sc i}, He\,{\sc
i} and coronal lines present during the eruption are present in all our
observations. From the fits of the computed infrared spectral energy 
distributions to the observed fluxes we find $T_{\rm eff}=4200\pm200$\,K 
for the red giant. 
The first overtone CO bands at 2.3\mic, formed in the atmosphere of the
red giant, are variable. The spectra clearly exhibit an infrared
excess due to dust emission longward of 5\mic; we estimate an effective
temperature for the emitting dust shell of 500~K, and find that the dust
emission is also variable, being beyond the limit of detection in
2007. Most likely, the secondary star in RS~Oph is intrinsically variable. 
\end{abstract}


\vskip2mm

\section{Introduction}

RS~Ophiuchi is a recurrent nova (RN) with at least six recorded outbursts, in
1898, 1933, 1958, 1967, 1985, and 2006 \citep{wallerstein}. The system consists
of a massive white dwarf (WD) and a red giant (RG) \citep{fekel}. Like
classical novae, eruptions are caused by a thermonuclear runaway (TNR) in
material  accreted on the surface of the WD \citep{starrfield}, although it is
unclear if the accretion disk is fed by Roche lobe overflow or the RG wind 
\citep{Starrfield08}.

RNe are divided into subclasses, depending on the nature of the secondary star
\citep{webbink}. During eruptions in RNe with a RG secondary (like RS~Oph), the
ejected material runs into the RG wind, which is shocked. Broad emission lines
arise in the shocked wind and ejecta, which narrow with time as the ejecta
decelerate \citep{das,evans07a}. In outburst, the infrared (IR) spectrum is
dominated by these lines, and by free-free radiation \citep{evans07b}. In
quiesence, the IR spectrum is dominated by the RG. 

Dust emission was detected within three years of the 1967 eruption by
\cite{geisel} and in 1983 by the InfraRed Astronomical Satellite (IRAS) survey
\citep{schaefer86}. More recently, \cite{evans07b} detected silicate dust
features $\sim7$ months after the 2006 outburst; they concluded that the dust
survives from one eruption to the next, and that some of the RG wind is
shielded from the shock and ultraviolet blast from the outbursts.

Knowledge of the elemental abundances in the atmosphere of the RG is important
for understanding the eruption, because the TNR on the WD occurs in material
accreted from the RG secondary. \cite{pavlenko08} modelled a 2006 August
spectrum of RS~Oph in the $1.4-2.5$\mic\ range and determined the following
parameters for the RG: $\rm T_{eff}=4100\pm100$\,K, $\log{g}=0.0\pm0.5$,
$\rm[Fe/H]=0.0\pm0.5, [C/H]=-0.8\pm0.2$ and $\rm[N/H]=+0.6\pm0.3$. These
abundances may vary considerably however, if the RG is contaminated by the nova
ejecta, as has been suggested by \cite{scott}. Irradiation of the RG by the
still-hot WD may also be a complicating factor in the immediate aftermath of an
eruption.

The latest outburst of RS~Oph was discovered on 2006 February 12.85~UT
\citep{hirosawa}; we assume the eruption began on this date ($t=0$). IR spectra
from the first 100 days of the outburst are discussed by
\cite{evans07a,evans07b}, \cite{das} and \cite{banerjee}.
 
Here we present IR spectroscopy of RS~Oph obtained on later dates and at
different orbital configurations. We investigate the effects of irradiation on
the secondary and dust emission at longer wavelengths.

\section{Observational data} 

An observing log is given in Table~\ref{obs}.
Figure~\ref{_vis} shows the times of our observations with respect to the visual light curve of RS~Oph.

\subsection{UKIRT}
IR spectroscopy of RS~Oph was obtained at the United Kingdom Infrared Telescope
(UKIRT), with the UKIRT $1-5$\mic\ Imager
Spectrometer (UIST) \citep{ramsay}. 
The observations were obtained in stare-nod-along-slit mode, with a 2-pixel-wide
slit. We obtained data in the $I\!J$ ($0.86-1.42$\mic), $H\!K$
($1.40-2.51$\mic), short $L$ ($2.91-3.64$\mic), long $L$
($3.62-4.23$\mic), and $M$ ($4.38-5.31$\mic) band grisms, giving a spectral
coverage of $0.86-5.31$\mic. The resolving power is $\sim600-2000$. We
obtained spectra of HR~6493 (F3\,V) immediately before we obtained spectra of
RS~Oph, for calibration purposes.

The data reduction method followed the usual routines for IR spectra.
The data at one nod position were subtracted from data at a second nod position
to remove sky emission lines, and the sky-subtracted data were extracted from
the UIST array. The extracted spectra of RS~Oph were divided by the extracted
spectra of the standard star to remove telluric absorption. The target was
observed at a similar airmass as the standard star to optimize the
cancellation of these features. The ratioed data were then multiplied by a
normalised blackbody to provide flux calibrated spectra; this assumes the
spectrum of HR~6493 is a 6700\,K blackbody, with $K=3.6$ \citep{tokunaga}.  
The data were wavelength calibrated using the spectra of arc lamps and telluric
lines in the standard star data.

\subsection{IRTF}

Observations were made using SpeX \citep{rayner} on the NASA Infrared
Telescope Facility (IRTF)\footnote{Data obtained by visiting astronomers at the 
Infrared Telescope Facility, which is operated by the University of Hawaii under 
Cooperative Agreement no. NCC 5-538 with the National Aeronautics and Space 
Administration, Science Mission Directorate, Planetary Astronomy Program.} in double-beam
mode. The slit dimensions were $0.8''\times15''$ and the nod was $7''$ along
the slit. The resolving power is $R\sim700-900$. To minimize atmospheric dispersion, the parallactic angle was set
such that the slit was oriented vertically. Internal wavelength calibration and
standard star observations were interspersed between spectra of RS Oph. In all
cases the star HD159170 (A5V) was used as a standard, and was within 0.07
airmass of RS Oph. All data were reduced with SpeXTools \citep{cushing}. To
generate the flux model for the standard star, we took the spectral type and
measured colour $(B-V)$, then scaled Kurucz' model spectra
\citep{kurucz94,kurucz95} to the $V$ flux.  

\subsection{Orbital phase}
The orbital phase of RS~Oph ($\Phi$, defined such that maximum radial velocity
of the RG occurs at $\Phi=0.00$) at the time of our observations is shown in
Table~\ref{obs} (phase at $t=0$ is $\Phi=0.95$), and was calculated from the orbital period and a zero phase
reference \citep{fekel}. In Table~\ref{obs}, we also show the visible fraction
of the RG hemisphere irradiated by the WD at the time of the
observations, assuming inclination $50^\circ$ (\citealt{Brandi09} derive $i=49-52^\circ$; \citealt{dobrzycka-k}, $i=30-40^{\circ}$; and \citealt{ribeiro}, $i=39\pm^1_{10}$).

The spectra presented in this paper are dereddened using
$E(B-V)=0.73$ \citep{snijders}. 

\begin{table*}
\centering
\caption[Observations of RS~Oph]{Observing log, best-fitting parameters,
and CO $\upsilon=2\rightarrow0$ band depth as a percentage of the continuum.}
\label{obs}
\begin{tabular}{lcccccccc}

\hline
\multicolumn{1}{c}{UT Date}&
\multicolumn{1}{c}{JD$^a$}&
\multicolumn{1}{c}{$t$ (days)$^b$}&
\multicolumn{1}{c}{Telescope}&
\multicolumn{1}{c}{$\Phi^c$} &
\multicolumn{1}{c}{$f^{d}$} &
\multicolumn{1}{c}{$T_{\rm eff}$ (K)}&
\multicolumn{1}{c}{$T_{\rm d}$ (K)} &
\multicolumn{1}{c}{CO \%}\\

\multicolumn{1}{c}{}&
\multicolumn{1}{c}{}&
\multicolumn{1}{c}{}&
\multicolumn{1}{c}{}&
\multicolumn{1}{c}{} &
\multicolumn{1}{c}{} &
\multicolumn{1}{c}{}&
\multicolumn{1}{c}{} &
\multicolumn{1}{c}{}\\

\multicolumn{1}{c}{(1)}&
\multicolumn{1}{c}{(2)}&
\multicolumn{1}{c}{(3)}&
\multicolumn{1}{c}{(4)}&
\multicolumn{1}{c}{(5)} &
\multicolumn{1}{c}{(6)} &
\multicolumn{1}{c}{(7)}&
\multicolumn{1}{c}{(8)} &
\multicolumn{1}{c}{(9)}\\

\hline

\multicolumn{1}{l}{2006 August 1}&
\multicolumn{1}{c}{3948}&
\multicolumn{1}{c}{170}&
\multicolumn{1}{c}{IRTF}&
\multicolumn{1}{c}{0.33}&
\multicolumn{1}{c}{0.16}& 
\multicolumn{1}{c}{4200}&
\multicolumn{1}{c}{500}&
\multicolumn{1}{c}{35}\\

\multicolumn{1}{l}{2006 August 9}&
\multicolumn{1}{c}{3956}&
\multicolumn{1}{c}{178}&
\multicolumn{1}{c}{UKIRT}&
\multicolumn{1}{c}{0.34}&
\multicolumn{1}{c}{0.18}&
\multicolumn{1}{c}{4200}&
\multicolumn{1}{c}{500}&
\multicolumn{1}{c}{34}\\

\multicolumn{1}{l}{2006 August 25$^e$}&
\multicolumn{1}{c}{3972}&
\multicolumn{1}{c}{194}&
\multicolumn{1}{c}{UKIRT}&
\multicolumn{1}{c}{0.38}&
\multicolumn{1}{c}{0.26}&
\multicolumn{1}{c}{4200}&
\multicolumn{1}{c}{500}&
\multicolumn{1}{c}{27}\\

\multicolumn{1}{l}{2006 September 18$^{e,f}$}&
\multicolumn{1}{c}{3996}&
\multicolumn{1}{c}{218}&
\multicolumn{1}{c}{UKIRT}&
\multicolumn{1}{c}{0.45}&
\multicolumn{1}{c}{0.40}&
\multicolumn{1}{c}{4200}&
\multicolumn{1}{c}{400}&
\multicolumn{1}{c}{31}\\

\multicolumn{1}{l}{2007 June 1}&
\multicolumn{1}{c}{4252}&
\multicolumn{1}{c}{474}&
\multicolumn{1}{c}{IRTF}&
\multicolumn{1}{c}{0.99}&
\multicolumn{1}{c}{0.52}&
\multicolumn{1}{c}{4400}&
\multicolumn{1}{c}{---}&
\multicolumn{1}{c}{26}\\

\multicolumn{1}{l}{2007 June 11}&
\multicolumn{1}{c}{4262}&
\multicolumn{1}{c}{484}&
\multicolumn{1}{c}{UKIRT}&
\multicolumn{1}{c}{0.01}&
\multicolumn{1}{c}{0.48}&
\multicolumn{1}{c}{4200}&
\multicolumn{1}{c}{---}&
\multicolumn{1}{c}{22}\\

\multicolumn{1}{l}{2007 July 28}&
\multicolumn{1}{c}{4309}&
\multicolumn{1}{c}{531}&
\multicolumn{1}{c}{IRTF}&
\multicolumn{1}{c}{0.12}&
\multicolumn{1}{c}{0.26}&
\multicolumn{1}{c}{4400}&
\multicolumn{1}{c}{---}&
\multicolumn{1}{c}{20}\\

\multicolumn{1}{l}{2008 July 15}&
\multicolumn{1}{c}{4662}&
\multicolumn{1}{c}{884}&
\multicolumn{1}{c}{UKIRT}&
\multicolumn{1}{c}{0.89}&
\multicolumn{1}{c}{0.72}&
\multicolumn{1}{c}{4000}&
\multicolumn{1}{c}{500}&
\multicolumn{1}{c}{27}\\

\hline

\multicolumn{9}{l}{$^a$ Julian Date -- 2450000.} \\
\multicolumn{9}{l}{$^b$ $t=0$ is 2006 February 12.85} \\
\multicolumn{9}{l}{$^c$ Orbital phase calculated from \cite{fekel}.} \\ 
\multicolumn{9}{l}{$^d$ Fraction of visible RG surface irradiated by WD.}\\
\multicolumn{9}{l}{$^e$ Data in \cite{pavlenko08}.} \\
\multicolumn{9}{l}{$^f$ IJ to $\rm L'$ on this date, M on 2006 September 19.} \\

\end{tabular}
\end{table*}

\begin{figure} 
\centering 
\includegraphics[width=80mm,angle=0]{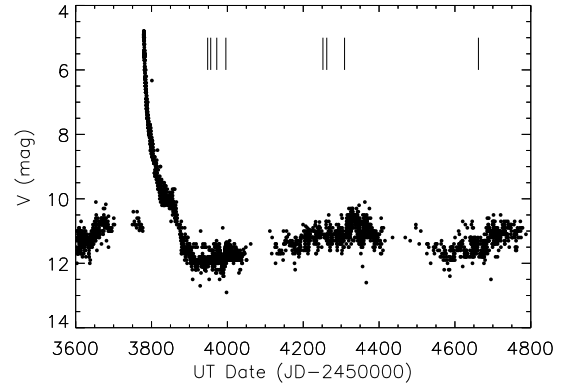} 
\caption{The visual light curve of RS~Oph from the American Association of Variable Star Observers (AAVSO), covering the 2006 outburst. The vertical lines show the dates of our observations.}
\label{_vis} 
\end{figure} 

\begin{figure} 
\centering 
\includegraphics[width=90mm,angle=0]{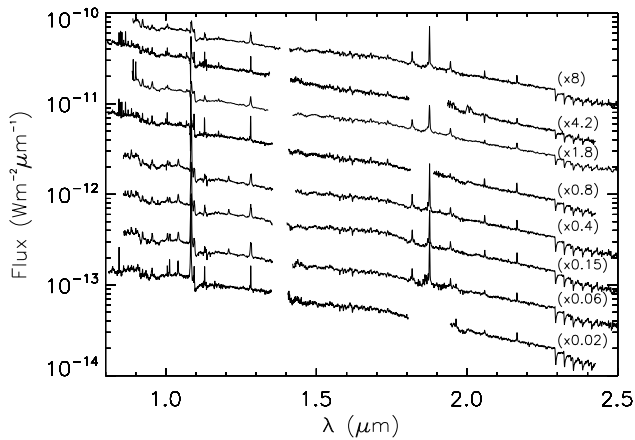} 
\caption{Near-IR spectra of RS~Oph in the $0.8-2.5$\mic\ range. From bottom to
top, the observing dates (UT) are 2006 August 1, 2006 August 9, 2006 August 25, 2006
September 18, 2007 June 1, 2007 June 11, 2007 July 28, and 2008 July 15. The
data have been multiplied by the amounts in brackets to vertically offset for
clarity. The spectra have been dereddened using $E(B-V)=0.73$ \citep{snijders}.}
\label{_ident} 
\end{figure}

\begin{figure} 
\centering 
\includegraphics[width=80mm]{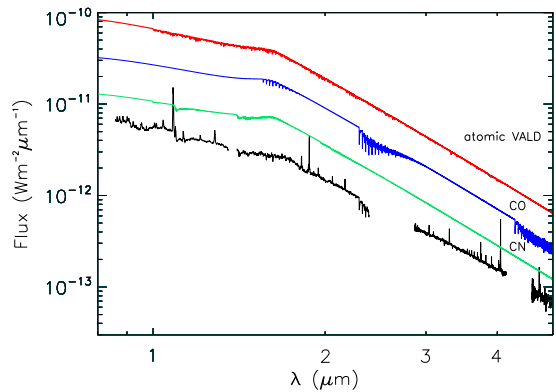} 
\caption{  
Computed spectra showing the $^{12}$CO, $^{13}$CO, CN, and atomic contributions
to the observed data (bottom spectrum). The computed spectra are shifted
vertically to aid comparison. The emission lines following the eruption are
still present in RS~Oph.} 
\label{_inent} 
\end{figure} 

\section{Results}

\subsection{Model fits to observed spectra}

IR spectra of RS~Oph in the range 0.9--2.4\mic\ are shown in
Figure~\ref{_ident}. They contain the H\,{\sc i}, He\,{\sc i}, and coronal
emission lines reported in earlier spectra by \cite{evans07a,evans07b} and Evans
et al. (in preparation). These lines 
originate in the shocked wind and gradually fade from the data. \cite{evans07a}
report early 2006 IR spectra, when the continuum is dominated by emission from
the hot gas. The earliest spectra reported here are from 2006 August. By this
time, the emission from the gas had subsided and the RG dominates the IR 
spectrum. Besides emission lines from the shock, the spectra in
Figure~\ref{_ident} contain absorption features from the secondary; these are
the subject of this paper.

Photometry of RS~Oph shows that the $V$ flux declined below the pre-outburst
value and reached a minimum $\sim200$~days after maximum. The $V$ flux rose
thereafter, reverting to the quiescent level after about 400~days
\citep{darnley}. IR photometry during eruption has been published for the 1985
outburst \citep{evans88} and (for a more limited time-base) by \cite{banerjee}.
These data show that the behaviour in the IR mirrors that in the optical. In our
data, the pre-outburst flux is essentially recovered by $t=474$~days.

The IR continuum and absorption spectrum of RS~Oph have been analysed with a
spectral synthesis technique, using  $\log{g}=0.0\pm0.5$, 
$\rm[Fe/H]=0.0\pm0.5, [C/H]=-0.8\pm0.2$, and $\rm[N/H]=+0.6\pm0.3$;
\citep[see][ for more details]{pavlenko08}, and microturbulent velocity 
$\rm V_t$ = 3\,km\,s$^{-1}$ \citep{pavlenko09}. Contributions of the
various molecular and atomic species to the total opacity are shown in
Figure~\ref{_inent}. The strongest molecular absorption is from the CO first
overtone ($\Delta\upsilon=2$, where $\upsilon$ is the vibrational quantum
number) and the CN red system ($A^2\Pi-X^2\Sigma^+$). 
As well as these features, a broad emission peak at 1.6\mic, which is
observed in late type stars, is present in RS~Oph and is due to the minimum
in the H$^-$ bound-free and free-free opacity \citep{john}.

The best-fitting model parameters are given in columns (7) and (8) of
Table~\ref{obs} for the dates of observation. Most fits were obtained with 
the best-fitting parameter $T_{\rm eff}=4200 \pm 200$\,K.  
The values $T_{\rm eff}$ in Table~\ref{obs} are essentially the same within the
uncertainties.

\begin{figure} 
\centering 
\includegraphics[width=80mm,angle=0]{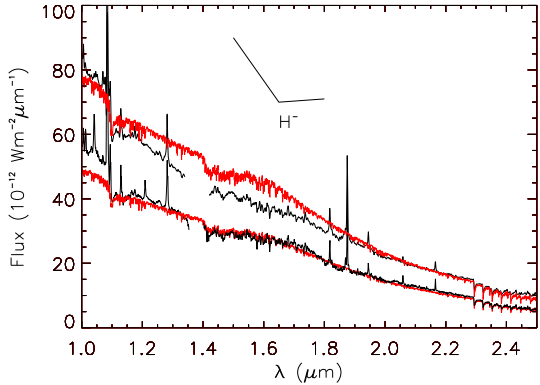} 
\caption{Best-fitting models ({\it red}) to the RS~Oph data on two dates ({\it
black}); the bottom spectrum is from 2006 August 25  and the top spectrum is
from 2007 June 11. A large discrepancy between the model and data is apparent
in the $1.2-2.0$\mic\ range on 2007 June 11. The wavelength range of the
minimum in the continuous absorption coefficient of the negative hydrogen ion is
shown above the spectra.}
\label{_fits} 
\end{figure}

We note that the model is unable to fit all parts of the data. The largest
discrepancy is seen in 2007, around the H$^-$ opacity minimum.
This problem is apparent in all data, but it is more acute in 2007 (see
Figure~\ref{_fits}). The peak due to the H$^-$ minimum is absent from these
spectra, implying a change in the IR spectrum of the
RG.

\begin{figure} 
\centering 
\includegraphics[width=80mm,angle=0]{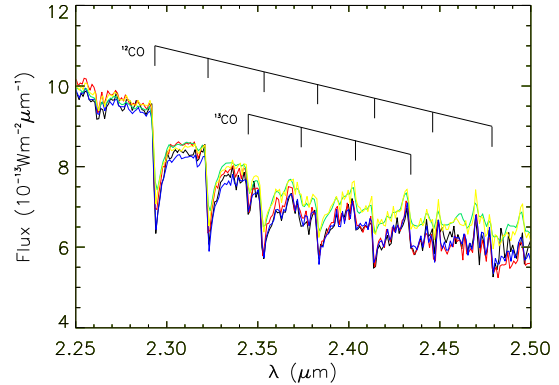} 
\caption{CO first overtone bands in RS~Oph at $R=2000$. Observation dates (UT) are 2006 August 9
({\it black}), 2006 August 25 ({\it red}), 2006 September 18 ({\it blue}), 2007
June 11 ({\it green}), and 2008 July 15 ({\it yellow}). The spectra have been
vertically shifted to the level of the 2006 August 9 spectrum to aid comparison.
The vertical lines show the positions of the $^{12}$CO and $^{13}$CO vibrational
band heads. The CO absorption is weaker on 2007 June 11 than it is on the other
dates.}  
\label{co} 
\end{figure}

Further evidence of a change in the RG can be seen in Figure~\ref{co}. In this
figure, the CO first overtone bands are shown in all our UKIRT data. Bands from
$^{13}$CO, as well as the main isotopomer, $^{12}$CO, are present \citep[the
isotopic ratio $\rm^{12}C/^{13}C$ for the RG is
$\rm^{12}C/^{13}C=22\pm3$;][]{pavlenko09}. The CO absorption is clearly weaker in
the 2007 data than it is in the other data. The CN bands may also be weaker, but the change is less pronounced than it is in the CO. The strength of the CO
$\upsilon=2\rightarrow0$ band is shown in column~(9) of Table~\ref{obs} as a
percentage of the continuum, from which it can be seen that the absorption is
weakest on 2007 July 28. Possible reasons for the behaviour of the CO are given below.

\subsection{The dust continuum}
\label{DC}
\begin{figure} 
\centering 
\includegraphics[width=80mm,]{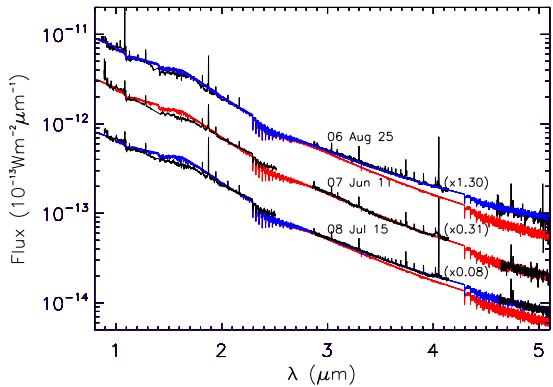} 
\caption{Model fits to the observed data (black lines) over the $0.8-5$\mic\
range. From top to bottom, the observation dates (UT) are 2006 August 25, 2007
July 11, and 2008 July 15. The red lines are model atmosphere fits,
the blue lines are fits obtained by the addition of a 500\,K blackbody 
to the model atmosphere. The flux in the model drops below the flux in the
data in the $\lambda>3$\mic\ region in the 2006 and 2008 data. This IR excess is
due to emission by dust. The data have been multiplied by the amounts in brackets to vertically offset for
clarity.}
\label{dust} 
\end{figure}

Figure~\ref{dust} shows the best-fitting model to the data on 2006 August 25, 2007 July 11 and 2008 July 15
over the $0.8-5$\mic\ range. The flux in the model falls below the flux in the
data for wavelengths \gtsimeq3\mic: RS~Oph clearly exhibits an IR excess. There
are two possible explanations for this excess: free-free radiation and dust
emission.

\cite{evans07b} reported the appearance of the silicate features at 9.7\mic\
and 18\mic\ in {\it Spitzer} data between 2006 April and 2006 September. They
concluded the dust was present before the outburst, that hot gas emission masked the
dust emission in the earlier data, but as the emission from the gas subsided,
the silicate features were revealed. Since free-free emission had faded by late
2006, and the silicate features were observed at this time, dust is the likely
source of the excess in our data.

\cite{vanloon} used the {\sc dusty} code to model the {\it Spitzer} spectrum and
found a dust temperature of 600\,K. From his fit, we find the dust contributes
$\sim7\times10^{-14}\,\rm W\,m^{-2}\mic^{-1}$ at 4\mic, approximately 50\%
of the flux in our 2006 data, and 40\% in our 2008 data. 

Since the contribution is significant at the longest wavelengths in our
data, and the dust emission peaks longward of $\sim5$\mic, we use a blackbody in 
our computations to mimic the dust emission: the effect of a $\nu^{\beta}$
dust emissivity law on the Wein tail is not expected to be significant.
The inclusion of a blackbody greatly improves our fits to the $3-5$\mic\ region
for all but three spectra: 2007 June 1, 2007 June 11 and 2007 July 28; the dust emission is at least a factor of 3 weaker in 2007.
Figure~\ref{dust} shows the dust excess on 2006 August 25, its absence on 2007 June 11, and its presence again on 2008 July 15.

The dust temperatures ($T_{\rm d}$) are given in column (8) of Table~\ref{obs}.
The uncertainty in $T_{\rm d}$ is at least $\pm100$\,K because of the absence
of data at $\gtsimeq5$\mic, where the dust emits most strongly and dominates
the continuum. The temperature of the dust is consistent with that in
\cite{vanloon}, and the dust contribution is consistent with an extrapolation
of the {\it Spitzer} fluxes. Therefore, we are confident that the excess we see
is emission from the dust detected by \cite{evans07b}. These observations
therefore support the claim that the dust will be present at the next eruption
\citep{evans07b}.  

As already noted, the 2007 spectra are unusual and are further discussed below;
we note that these data were obtained at both UKIRT and IRTF, so there is no
doubt that the effect is real and not instrumental or an artefact of data
reduction. 

\section{Discussion and conclusions}

\subsection{CO}

In this paper we have presented IR spectroscopy of RS~Oph obtained
on eight occasions after the 2006 outburst, starting on $t=170$~days and ending
on $t=884$~days. Although the values of effective temperatures on all observing
dates are generally consistent, we see changes in the spectral features from
the RG, the most obvious being a significant weakening of the CO first overtone
bands. We now consider possible explanations for this behaviour. 

The weakening of the CO first overtone bands in RS~Oph has been observed
before, following the 1985 eruption. \cite{scott} obtained IR spectroscopy of
RS~Oph in quiesence in 1992. The strength of the CO absorption features had
declined significantly from an observation seven years earlier, 143~days after
the 1985 outburst \citep{evans88}. \cite{harrison} obtained IR spectra of
RS~Oph in 1992, approximately one month from \citeauthor{scott} The CO was
absent from their data, but the signal-to-noise ratio was low and weak
absorption could have been present. \cite{harrison} concluded that CO emission
from an accretion disc veiled the photospheric features of the RG. An
alternative explanation was given by \cite{scott}: they concluded the outburst
contaminated the secondary with carbon, which was then convected away,
restoring the carbon abundance to its original value.

\cite{scott} dismissed the \citeauthor{harrison} explanation because of the
absence of CO emission in high signal-to-noise data. The same conclusion can be
reached from our data, which show only band absorption profiles consistent with
a late-type star. \citeauthor{harrison} based their conclusion on the
discovery  of CO first overtone emission in the accretion discs around pre-main
sequence stars. However, CO first overtone emission has been detected in
few eruptive variables: six classical novae (NQ~Vul, V842~Cen,
V705~Cas, V1419~Aql, V2274~Cyg and V2615~Oph), in which the CO formed in cool, neutral
regions in the ejecta \citep{Ferland79,Hyland89,Lynch95,Evans96,Rudy03,Das09}
and the peculiar eruptive variable V838 Mon \citep{Rushton05}. We do not 
therefore expect CO emission in RS~Oph, as the temperature of the shocked gas
is $\sim10^6$\,K \citep{evans07a}.

According to \cite{scott}, the CO bands deepen when the nova ejecta contaminate
and sweep past the secondary.  \cite{Drake09} and \cite{Ness09} argue that the
C is overabundant in the nova ejecta from an analysis of {\it Chandra} X-ray
data. Since the C-enriched ejecta would pollute the RG within hours of the
outburst,  and any carbon deposited in the atmosphere would quickly combine
with oxygen to form CO, \citeauthor{scott} suggest that the CO bands deepen
shortly after optical maximum. They then argue that the CO weakens, as
convection reduces the carbon abundance to the pre-outburst value, which is
reached after $\sim10^2$ days. However, the CO absorption weakened and then
deepened in our data, inconsistent with the behaviour predicted by
\citeauthor{scott}.  Therefore, contamination of the secondary alone cannot
explain the evolution of the CO.

The weakening in our data of the CO absorption and the disappearance of the
H$^-$ opacity minimum peak imply a higher RG temperature in 2007. A possible
explanation is that the WD heats the RG hemisphere facing the primary. As the
star orbits the WD, varying fractions of the irradiated hemisphere of the RG
are presented to the observer and the depths of the molecular bands change with
phase $\Phi$. The problem with this explanation is that the weakest CO is
observed at $\Phi=0.12$, when 26\% of the irradiated hemisphere is
visible, and not at other phases, when similar or even larger fractions are
visible. We would expect the CO absorption to be weakest when we see the
largest fraction of the irradiated hemisphere and the effect of irradiation
(and hence destruction of CO) is at its maximum.

\cite{anupama} monitored RS~Oph in quiesence and observed variability in the
absorption from the RG. From the [TiO]$_1$, [TiO]$_2$, [VO] and [Na] spectral
indices (defined by \citealt{kenyon}), they found that features at  shorter
wavelengths imply an earlier spectral type than features at longer wavelengths
\citep[see also Table~5 in][]{dobrzycka}. They concluded that the variable blue
continuum from the hot component veiled the absorption from the RG. This effect
was noticed in other symbiotic stars by \cite{kenyon}, who showed that the
spectral indices are correlated  with the visual brightness of the system.
Although this explains variability of optical absorption features, it is
unlikely to be responsible for the behaviour of the CO, as the contribution
from the hot component is negligible at 2.3\mic. Furthermore, the CN bands,
which are in the blue part of the spectrum, would show a significant and larger
change, but this is not observed in our data.

The only alternative explanation for the behaviour of the CO is that the RG is
intrinsically variable. \cite{rosino} estimated as M2-III the spectral type of
the RG. They had monitored RS~Oph for 12 years after the 1967 outburst and
noted that the secondary is ``perhaps slightly variable'', although the
variability they observed could be due to the blue continuum, as mentioned
above.

More pertinent to the subject matter of this paper is the measurement by
\cite{kenyon-co} of a ``CO index'' ($[\rm CO] = [2.4]-[2.17]$, where $[2.4]$
and $[2.17]$ are narrow-band magnitudes in the region of the CO first overtone
bandhead) in 1985--1986, in the aftermath of the 1985 eruption; 
\citeauthor{kenyon-co} found the CO index to be variable. It is unfortunate
that, apart from \cite{scott} and  \cite{harrison}, there has been no $1-2.5\,\mu$m
spectroscopy of RS~Oph in quiescence. The spectroscopic variability of the RG
should be confirmed by regular IR monitoring, where the contribution from the
hot component is negligible.  

\subsection{Dust}

In this paper we show that RS~Oph exhibits an excess at $>3$\mic\ (see
Figure~\ref{dust}) due to emission from the dust already known to exist in the
system. The dust is present in quiesence and survives the UV flash of the
outburst. \cite{evans07b} suggest that the dust is largely confined to the
binary plane and that this higher density material is effectively shielded from
the eruption. Furthermore, if the dust temperature is $T_{\rm d}=500$\,K, the
dust is heated in outburst to only $T_{\rm d}=1250$\,K, below the sublimation
temperature, although the situation is marginal \citep[see][]{evans07b}. This
dust will then be present at the next eruption, provided it survives the
passage of the shock.

The fitting analysis (see Section~\ref{DC} and Table~\ref{obs}) seems to
imply that the dust may not have survived, as the IR excess seems to have
disappeared between 2006 September and 2007 June, only to develop again,
shortly after. Figure~\ref{_vis} shows that the 2007 spectra (showing no dust excess) were obtained during a small-amplitude rebrightening event in the visual light curve, while the earlier and later spectra (with the dust excess) were obtained when the visual magnitude was close to the quiescent value. Since the 2007 spectra show no IR excess, the rise in the $V$ band flux might be interpreted in terms of the dissipation of the dust and its subsequent decline to the formation of new dust in the cooling ejecta. However, this interpretation of the light curve is unlikely, as the optical depth in the visual due to the pre-outburst dust is only $\tau_{\rm V}=0.1$ \citep{vanloon}. Furthermore, the temperature of the newly formed silicate dust would be close to the condensation temperature ($\simeq1300$\,K; \citealt{Speck00}), much higher than the temperature we find after the redevelopment of the excess, $\sim500$\,K. Also, as discussed in \cite{evans07b}, conditions are unlikely to be suitable for dust condensation in the ejecta. 
Moreover, as our data only go  as far as 5\mic, it is in any case possible
that an IR excess is still present in 2007, but at wavelengths longer than are covered
by our data. 

However, it is interesting that the disappearance of the IR excess shortward of
5\mic\ coincides almost exactly with a change in  the 9.7\mic\ silicate feature
in {\it Spitzer} spectra \citep{evans07b}: the narrower feature is present in
2007 April, $\sim2$ months before we see no excess. We will present a detailed
discussion of the circumstellar dust in the RS~Oph system in a separate paper. 

\section{Summary}

We have presented IR spectroscopy of the RN RS~Oph on eight occasions 
after the most recent outburst in 2006 February. The spectra contain emission 
lines from the outburst superimposed on the spectrum of the RG in the system. We have fitted synthetic spectra to the data to determine the 
effective temperature of the RG and dusty envelope. 
Although the parameters on all observation dates are consistent within 
the uncertainties, the spectral features from the RG are variable. This variability cannot be explained by contamination of the RG, irradiation, or veiling from the hot component. 
The most likely explanation is that the RG is intrinsically variable. 
This variability should be confirmed by monitoring of RS~Oph in the IR, where the RG dominates the continuum.

The spectra show an IR excess at $>3$\mic\ due to emission from the 
circumstellar dust detected in earlier studies. The excess is present in our 2006 and 2008 data, but absent in our 2007 data. However, it is possible the excess is present in 2007, but at longer wavelengths than are covered by our data. 

\vskip2mm

{\it Acknowledgement}

This work was supported in part by the Aerospace Corporation's Independent
Research and Development (IR\&D) program. The authors are grateful to John
Rayner with assistance with SpeX and Dagny Looper for exchanging some telescope
time with us.

The work was also supported by an International Joint Project Grant 
from the UK Royal Society and the ``Microcosmophysics'' program of 
the National Academy of Sciences and Space Agency of Ukraine. 
 
We acknowledge with thanks the variable star observations from the AAVSO international database contributed by observers worldwide and used in this research. We thank the anonymous referee for helpful comments.

\end{document}